\documentclass[lettersize,journal]{IEEEtran}
\usepackage{amsmath,amsfonts}
\usepackage{array}
\usepackage{textcomp}
\usepackage{stfloats}
\usepackage{url}
\usepackage{verbatim}
\usepackage{graphicx}
\usepackage{subcaption}
\usepackage{cite}
\graphicspath{{./Figure/}}
\usepackage{amssymb}
\usepackage{algorithm}
\usepackage{algpseudocode}
\floatname{algorithm}{Protocol}
\usepackage{float}
\usepackage{threeparttable}
\begin{document}

\title{Lifelong reinforcement learning for health-aware fast charging of lithium-ion batteries}

\author{Meng Yuan,~\IEEEmembership{Member,~IEEE,}, Changfu Zou,~\IEEEmembership{Senior Member,~IEEE}
        % <-this % stops a space
\thanks{This work was supported by the European Union (EU)-funded Marie Sklodowska-Curie Actions (MSCA) Postdoctoral Fellowship (Grant No. 101110832).}% <-this % stops a space
\thanks{Meng Yuan and Changfu Zou are with the Department of Electrical Engineering, Chalmers University of Technology, 412 96 Gothenburg, Sweden. (Emails: {\tt\small meng.yuan@ieee.org; changfu.zou@chalmers.se}}}

\markboth{IEEE Transactions on Transportation Electrification, June~2025}% Journal of \LaTeX\ Class Files,~Vol.~14, No.~8, August~2021
{Shell \MakeLowercase{\textit{et al.}}: A Sample Article Using IEEEtran.cls for IEEE Journals}


\maketitle

\begin{abstract}
Fast charging of lithium-ion batteries remains a critical bottleneck for widespread adoption of electric vehicles and stationary energy storage systems, as improperly designed fast charging can accelerate battery degradation and shorten lifespan. In this work, we address this challenge by proposing a health-aware fast charging strategy that explicitly balances charging speed and battery longevity across the entire service life. The key innovation lies in establishing a mapping between anode overpotential and the state of health (SoH) of battery, which is then used to constrain the terminal charging voltage in a twin delayed deep deterministic policy gradient (TD3) framework. By incorporating this SoH-dependent voltage constraint, our designed deep learning method mitigates side reactions and effectively extends battery life. To validate the proposed approach, a high-fidelity single particle model with electrolyte is implemented in the widely adopted PyBaMM simulation platform, capturing degradation phenomena at realistic scales. Comparative life-cycle simulations against conventional CC-CV, its variants, and constant current–constant overpotential methods show that the TD3-based controller reduces overall degradation while maintaining competitively fast charge times. These results demonstrate the practical viability of deep reinforcement learning for advanced battery management systems and pave the way for future explorations of health-aware, performance-optimized charging strategies.
\end{abstract}


\section{Introduction}

% Importance of Lib and challenges. 
\IEEEPARstart{L}{ithium-ion} batteries (LIBs) have emerged as the cornerstone of modern energy storage systems, powering applications ranging from portable electronics to electric vehicles (EVs) and grid-scale renewable energy integration \cite{zhang2023charging}. Despite their dominance, EVs continue to face two persistent challenges, limited driving range and prolonged charging durations, which remain critical barriers to their widespread adoption and the push for energy-efficient electrification \cite{goodenough2013li,ahmad2025techno}. The energy density, charge and discharge kinetics of LIBs are primarily determined by their electrochemical composition, including cathode and anode materials and electrolytes, as well as their structural design, such as particle morphology and electrode architecture \cite{blomgren2016development}. While advancements in battery chemistry continue to push theoretical limits of battery, sophisticated battery management systems (BMS) have shown promise in optimizing charging protocols to achieve fast charging without exceeding intrinsic material constraints \cite{rahimi2013battery}. Though effective in reducing charging time, unconstrained high-current charging accelerates capacity fade through mechanisms such as lithium plating and solid-electrolyte interphase (SEI) growth \cite{broussely2005main}. To address this dilemma, advanced control algorithms capable of dynamically balancing charging speed and longevity, through multi-physics-aware current modulation and degradation-predictive interventions, are urgently required to unlock the full potential of LIBs \cite{attia2020closed}.

% Discuss the battery modelling
Accurate battery modeling is essential for designing health-aware fast charging control algorithms. One widely used approach is the equivalent circuit model (ECM), which simulates voltage dynamics using simplified resistor-capacitor networks and offers computational efficiency ideal for real-time embedded systems \cite{li2023nonlinear, yuan2025robust}. However, without modeling the first principles underlying intercalation reactions and the diffusion process within battery cells, ECMs often struggle to capture locally distributed behavior, particularly those related to safety and aging. In contrast, electrochemical models, such as the Doyle-Fuller-Newman (DFN) model, explicitly describe ion transport and reaction kinetics across electrode microstructures, enabling physics-based predictions of degradation modes \cite{khalik2021parameter}. Yet, the significant computational burden of solving the coupled partial differential equations (PDEs) inherent in the DFN models limits their practical application in advanced control methods that also demand high computational power. To address these challenges, the single particle model (SPM) has emerged as a pragmatic simplification. By assuming a uniform electrolyte concentration, the SPM reduces PDEs to a set of ordinary differential equations (ODEs) while still capturing essential electrochemical behavior \cite{northrop2014efficient}. Recent work has extended the SPM to incorporate various capacity fade mechanisms and integrated it into a model predictive control (MPC) framework to optimize charging profiles and minimize intra-cycle capacity fade \cite{hwang2022model}. However, the conventional SPM neglects  electrolyte dynamics and leads to significant voltage prediction errors at high C-rates, which has motivated the development of extensions such as the single particle model with electrolyte (SPMe) for improved accuracy \cite{moura2016battery, marquis2019asymptotic}. % li2020comprehensive

Conventional lithium‐ion battery charging typically employs the constant current–constant voltage (CC-CV) method, where the battery is first charged at a fixed current until a set voltage is reached, and then held at constant voltage while the current gradually decreases. Although fast charging can be achieved by increasing the current or target voltage, such measures tend to accelerate battery degradation phenomena such as lithium plating, thereby reducing battery life \cite{huang2022onboard}. To solve this issue, researchers turned to health‐aware fast charging strategies that integrate battery degradation models with advanced control techniques. In \cite{wassiliadis2023model}, an anode overpotential based PID controller is designed to mitigate the risks of lithium plating while increasing the speed of charging. In \cite{yang2022constant}, Yang et al. proposed a constant-overpotential based fast charging strategy for Li-ion batteries and show that this constant-overpotential control outperforms the traditional CC-CV protocol in both charging speed and lithium-plating suppression. Similarly, Lu et al. developed a MPC framework that incorporates real-time lithium plating detection and adaptive parameter updates to optimize the balance between fast charging and long-term battery health \cite{lu2024health}. However, this model based approach is not adaptive to the parameters variation during battery degradation. Moreover, the overpotential is hard to be measured in practical application, which hinders the implementation of overpotential based method. This challenge has inspired the development of various approaches including machine learning-based methods, to estimate lithium plating \cite{zhang2025machine}.

To address the challenges of model mismatch and parameter drift as batteries age, learning-based methods have been introduced in the field of battery charging. In \cite{park2022deep}, a deep reinforcement learning (RL) based approach has been designed for the fast charging of battery, and two cases considering if the overpotential is measurable are discussed in the design of the controller. While the method shows promise for rapid charging, it models battery aging solely as an increase in film resistance, and the controller design does not integrate long-term battery degradation considerations. In \cite{wei2021deep}, Wei et al. propose a deep RL-based strategy for the fast charging of lithium-ion batteries, specifically targeting thermal safety and health-conscious charging. The multiphysics-related constraints are implicitly incorporated into the reward design of the RL. However, the agent is trained on a per-episode basis, with the battery model initialized at the beginning of each epoch. In \cite{hao2023adaptive}, an adaptive model-based RL strategy leverages Gaussian processes to capture the battery environment and enforce operational constraints during charging, but its primary focus is on minimizing charging time rather than extending battery life.

Based on the discussion above, our work addresses the challenge in fast charging by taking battery degradation into explicit account throughout the entire lifespan. We employ a twin delayed deep deterministic policy gradient (TD3) method for health-aware fast charging. To mitigate battery degradation, we first establish a mapping between the side-reaction overpotential and the charge cut-off voltage and then integrate it as health-related constraint into the training of our RL agent. The SPMe coupling degradation is employed as the battery environment to capture electrochemical behavior. The contributions of this work are summarized as follows:

\begin{itemize}
    \item Unlike previous approaches that focus solely on minimizing charging time while only implicitly incorporating constraints to extend battery life, this introduces the first explicit formulation of lifelong battery fast charging problem, aiming at reducing charging duration while extending longevity of battery.
    
    \item A mapping between the charge cut-off voltage and SoH is established using a constant current (CC)--constant overpotential (COP) approach. This mapping leverages the intrinsic relationship between the anode overpotential and battery degradation, and can be integrated into both CCCV-based charging and learning-based advanced charging, allowing directly incorporation of battery health into charging control.

    \item By integrating the obtained mapping between the voltage and SoH, a TD3-based charging strategy is designed to explicitly optimize the fast charging process.
    
    \item The proposed TD3-based strategy is trained and tested through comprehensive life-cycle simulations using the battery simulator Python Battery Mathematical Modeling (PyBaMM) \cite{sulzer2021python} with a SPMe-aging model. Superior performance is demonstrated over CC-CV charging and its variants.
        
\end{itemize}

The remainder of this paper is organized as follows. Section~\ref{sec:problem_RL} defines the health-aware fast charging problem. Section~\ref{sec:RL_method} presents the TD3-based reinforcement learning approach. Section~\ref{sec:results} reports modeling details, experimental setup, and charging results, including analysis of aging effects and lithium plating. Finally, Section~\ref{sec:conclusion} concludes the paper.

\section{Health-aware fast charging problem}\label{sec:problem_RL}

A battery is considered to have reached its end of life when its SoH falls to a specified threshold, such as 75-80\% for EV applications. Within this health-aware fast charging framework, we aim to minimize the charging time needed to reach the desired SoC while maximizing the number of equivalent full cycles (EFCs) of battery until the SOH reaches 80\%. The cost function for this problem is defined as:

\begin{equation}
		J(t) = w_{1} C_{\text{SoC}}(t) + w_{2} C_{\text{side}}(t),
    \end{equation}
where \(w_{1}\) and \(w_{2}\) are weighting factors. The first part of the cost is designed for SoC setpoint tracking, aiming to minimize charging duration. The cost function is defined as: 
    \begin{equation}
		C_{\text{SoC}}(t) = \left| \text{SoC}^* - \text{SoC}(t) \right|,
    \end{equation}
where \(\text{SoC}^*\) denotes the target state of charge, and \(\text{SoC}_{t}\) represents the current state of charge. Considering the lithium plating is the main degradation mechanism here, the degradation-related cost is defined as:
    \begin{equation}
        C_{\text{side}}(t) = \left| \eta_{\text{side}}(t) - \eta_{\min} \right|,
    \end{equation}
where \(\eta_{\min}\) is a predefined threshold overpotential, below which battery degradation becomes excessively rapid. We can interpret this threshold as follows: when the anode overpotential falls below this value, further reducing it to achieve faster charging is no longer beneficial, as the resulting acceleration in battery degradation outweighs any additional gains in charging speed. 

The constraint of this problem involves that the input current is limited by the battery charger as:
    \begin{equation}
		0 \leq I(t) \leq I_{\max}. \label{eq:I_constraint}
    \end{equation}

Then, the health-aware fast charging problem is formulated as
    \begin{align}
        \underset{u(t)}{\min} & \quad \int_{0}^{t_{e}} J(t) dt \nonumber \\
        \text{s.t.} & \quad \text{Battery dynamics:} \quad \dot{x}(t) = f(x(t),u(t)), \nonumber \\
        & \quad \text{Input constraints:} \quad \eqref{eq:I_constraint}, \nonumber \\
        & \quad \text{SoC}(t_{f}) = \text{SoC}^{*}, \quad \forall \text{ cycles}, \nonumber \\
        & \quad \text{SoH}(t_{e}) = \text{SoH}_{\text{end}}, \label{eq:problem_formulation}
    \end{align}
where $x$ denotes the state of battery dynamics, $u$ is the charging current serving as the system input, \(t_{f}\) is the charging completion time for each cycle, which can vary over the lifespan of battery, $t_{e}$ represents the end-of-life time of the battery, and \(\text{SoH}_{\text{end}}\) is the specified battery SoH at end-of-life, set to 80\% in this study.

This optimization problem is formulated to minimize the overall cost function over the battery lifetime from initial time 0 to its end-of-life time $t_{e}$. Each charging cycle within this period terminates at time $t_{f}$ when the SoC reaches the target SoC$^*$.

\subsection{Constant current constant overpotential based control}\label{sec:CC-COP}

To solve the fast charging problem as shown in \eqref{eq:problem_formulation}, one intuitive idea is to design the controller based on overpotential feedback. This constant current-constant overpotential (CC-COP) method mirrors the traditional CC-CV approach by initiating with a constant current phase, followed by dynamically regulating the anode potential to a small positive value using real-time feedback. This method has been shown to effectively reduce the risk of plating by maintaining the electrode overpotential at a safe, constant level after the initial current phase \cite{yang2022constant, zhang2025machine}. % yang2022constant
	
However, the CC-COP control in fast charging faces two significant limitations: Firstly, measuring electrode potentials directly in commercial batteries is challenging as it requires advanced sensors typically used only in laboratory settings, making it impractical for widespread application. Secondly, determining the appropriate overpotential threshold is not straightforward. While setting a high threshold can help reduce lithium plating, it may result in overly conservative charging corresponding to very long charging times. The need to finely tune the overpotential threshold adds complexity to real-world implementation, further complicating the adoption of CC-COP control strategies in commercial battery charging systems. 

\section{Reinforcement learning based fast charging algorithm}\label{sec:RL_method}

\begin{figure*}[h!]
    \centering
    \includegraphics[width=0.8\linewidth]{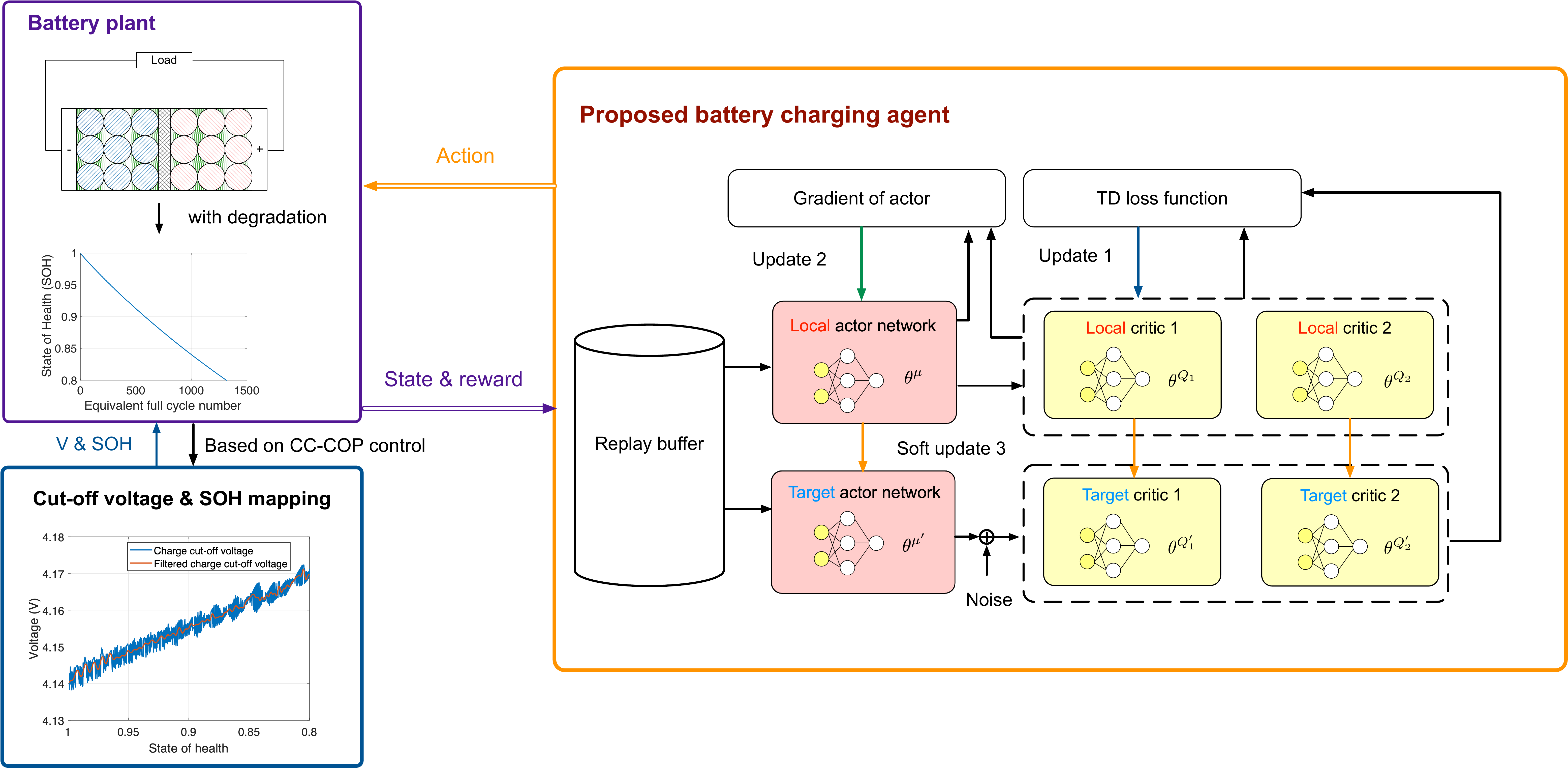}
    \caption{Structure of the proposed method.}
    \label{fig:entire_system}
\end{figure*}

To overcome the aforementioned limitations of CC-COP control, a data-driven alternative based on RL is adopted. Unlike model-based strategies that depend on the development and calibration of accurate physical models, RL learns optimal charging policies directly through interactions with the environment, using observable signals and reward feedback. This paradigm enables the development of adaptive charging strategies that can incorporate battery health indicators without requiring explicit physical modeling. 

In this work, we aim to design a controller that builds upon the insights provided by the CC-COP controller and introduce a twin delayed deep deterministic policy gradient (TD3)-based method to solve the health-aware fast charging problem defined in \eqref{eq:problem_formulation}. Fig.~\ref{fig:entire_system} illustrates the overall system structure, which comprises three main components. 

\begin{figure*}[htbp]
        \centering
        \includegraphics[width=0.6\linewidth]{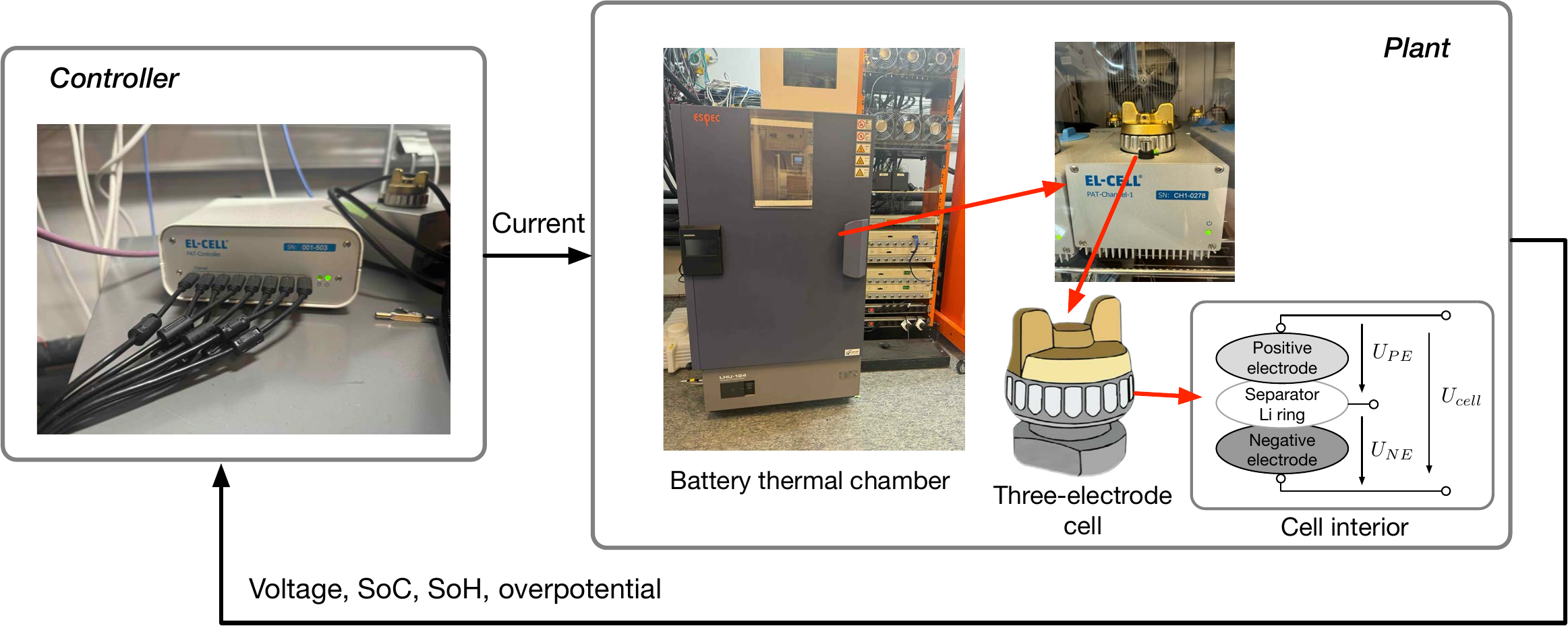}
        \caption{Experimental setup for CC-COP battery cycling.} 
        \label{fig:exp_setup}
    \end{figure*}

The first component is the battery environment, which may represent either a real battery system or a high-fidelity lithium-ion battery model that captures degradation dynamics. Since the anode overpotential is challenging to be measured in practical applications, we aim to acquire a correlation between the cut-off voltage and SoH based on the CC-COP controller, as detailed in Section~\ref{sec:CC-COP}. The maximum terminal voltage of the battery cell, $V_\text{cut-off}$, during each charging process that consistently aims for the SoC setpoint, $\text{SoC}^*$ is recorded during charging. This voltage identification continues until the SoH of battery falls below 80\%, signaling the end of its lifespan. This step can be experimentally performed using the setup shown in Fig.~\ref{fig:exp_setup}, where a PAT-Cell system from EL-CELL is employed for testing three-electrode cells, or via a high-fidelity simulation. The resulting correlation between charge cut-off voltage and SoH provides important battery health information to the environment, enabling the generation of reward signals for the battery charging agent. The third and core component of our system is the TD3-based charging agent, which receives state and reward information from the environment and computes optimal charging actions accordingly. The detailed formulation and implementation of this agent is presented in the following sections.

\begin{figure*}
    \centering
    \includegraphics[width=0.6\linewidth]{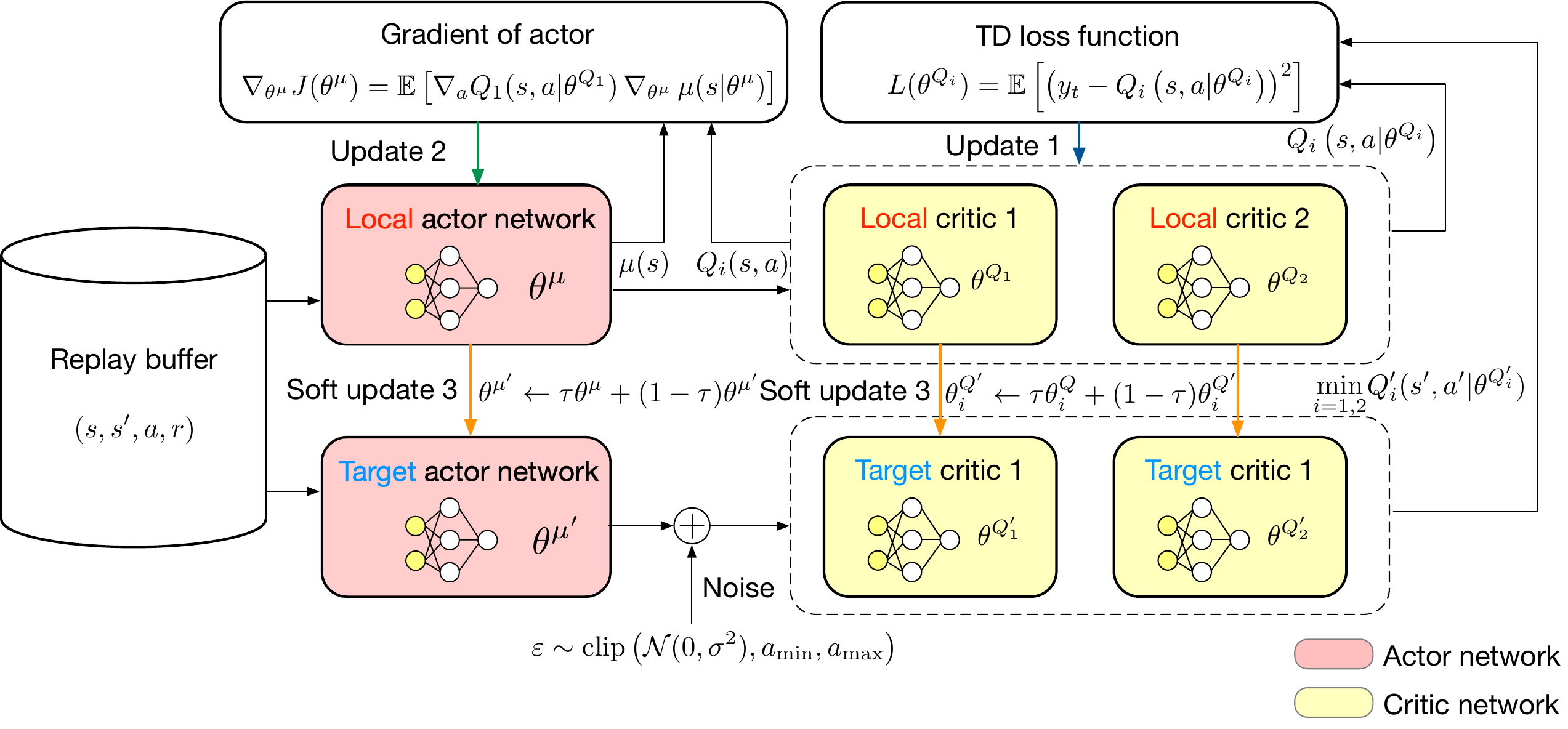}
    \caption{Structure of the deep learning algorithm.}
    \label{fig:TD3_structure}
\end{figure*}

\subsection{TD3 method overview}

To address the charging problem specified in \eqref{eq:problem_formulation}, the TD3 algorithm, an enhanced version of the DDPG framework, aims to rectify the Q-value overestimation issue inherent in DDPG. It introduces targeted improvements that boost the learning stability and accuracy. The architecture of the TD3 algorithm is illustrated in Fig.~\ref{fig:TD3_structure}. Within the Actor-Critic framework, TD3 incorporates two types of networks: the critic network $Q(s,a|\theta^{Q})$ with parameter $\theta^{Q}$ and the actor network $\mu(s|\theta^{\mu})$ with parameter $\theta^{\mu}$. To mitigate the overestimation of the Q-value, TD3 employs a technique that calculates the target Q-value by selecting the lesser value from the two target critic networks, effectively reducing potential overestimation:
	\begin{equation}
		y_{t}(r, s^{\prime}) = r(s,a) +\gamma \underset{i=1,2}{\min} Q^{\prime}_{i}(s^{\prime}, a^{\prime}| \theta^{Q^{\prime}_{i}}), 
	\end{equation}
	where $y_{t}$ is the target Q-value for given state $s$ and action $a$; $a^{\prime}$ is the next action chosen by the policy based on next state $s^{\prime}$; $\gamma$ is the discounting factor; $\theta^{Q^{\prime}_{i}}$ is the parameter of target critic. The critic network updates its parameters based on gradient descent applied to the loss function as follows: 
	\begin{equation}
		L(\theta^{Q_{i}}) = \mathbb{E} \left[ \left( y_{t} - Q_{i}\left(s, a | \theta^{Q_{i}} \right) \right)^2 \right], 
	\end{equation}
	\begin{equation}
		\nabla_{\theta^{Q_{i}}} L(\theta^{Q_{i}}) = \mathbb{E} \left[ \left( y_{t} - Q_{i}(s, a | \theta^{Q_i}) \right) \nabla_{\theta^{Q_i}} Q_{i}(s, a | \theta^{Q_i}) \right], 
	\end{equation}
	where \( \mathbb{E}(\cdot) \) denotes the expectation operator, $\theta^{Q_{i}}$ is the parameter of $Q_{i}$. As indicated in Fig.~\ref{fig:TD3_structure}, the parameter is updated (Update 1) according to:
	\begin{equation}
		\theta^{Q_i} \leftarrow \theta^{Q_i} - \alpha  \nabla_{\theta^{Q_{i}}} L(\theta^{Q_{i}}),
	\end{equation}
	where $\alpha$ is the learning rate of the critic network. Moreover, TD3 reduces instability by postponing updates to its actor network relative to its critic. In practice, the critic is updated frequently to accurately capture the dynamics of environment, while the actor is adjusted only after a fixed number of critic updates. This delay minimizes error propagation and keeps the policy changes smoother, resulting in more stable and reliable learning outcomes. The actor network is updated (Update 2) with parameters using gradient ascent of the objective function as: 
	\begin{equation}
		J_{Q}(\theta^\mu) = \mathbb{E}\left[ Q_1(s, \mu(s)) \right],
	\end{equation}
	\begin{equation}
		\nabla_{\theta^\mu} J_{Q}(\theta^\mu) 
		= \mathbb{E}\left[ \nabla_a Q_1(s, a | \theta^{Q_1})\,\nabla_{\theta^\mu}\,\mu(s | \theta^\mu)\right], 
	\end{equation}
	\begin{equation}
		\theta^\mu \leftarrow \theta^\mu \;+\; \beta\,\nabla_{\theta^\mu} J_{Q}(\theta^\mu), 
	\end{equation}
	where $\theta^{\mu}$ is the parameter of actor network, $J(\theta^{\mu})$ is the defined objective function, $\nabla_{\theta^\mu} J(\theta^\mu)$ is the gradient of objective function with respect to $\theta^{\mu}$, $\beta$ is the learning rate of the policy network. As denoted as Soft update 3 in the Fig.~\ref{fig:TD3_structure}, the parameters of the target critic and the target actor network are updated as: 
	\begin{gather}
		\theta^{Q^{\prime} }_{i} \leftarrow \tau \theta^{Q}_{i} + (1 - \tau) \theta^{Q^{\prime}}_{i},\\
		\theta^{\mu^{\prime}} \leftarrow \tau \theta^{\mu} + (1 - \tau) \theta^{\mu^{\prime}},
	\end{gather}
	where the soft updating factor is denoted by $\tau$ and $\theta^{\mu^{\prime}}$ is the parameter of the target actor network.
	
	Furthermore, TD3 introduces truncated normal noise to the actions produced by the target policy network, mitigating the trade-off between bias and variance. This added noise serves as a form of regularization, which helps prevent overestimation of Q-values and reduces the risk of overfitting. By smoothing the target updates in this manner, TD3 achieves more stable learning and enhances the reliability of policy evaluation:
	\begin{equation}
		a_{t+1} \leftarrow \mu^{\prime}(s^{\prime}|\theta^{\mu^{\prime}}) + \varepsilon, \, \varepsilon \sim \text{clip} \left( \mathcal{N}(0, \sigma^2), a_{\min}, a_{\max} \right),
	\end{equation}
	where $a_{\min}$ and $a_{\max}$ define the valid range of actions, representing the lower and upper bounds of the battery charging current, respectively.

\subsubsection{Battery application}

For the TD3-based fast charging algorithm, the battery SoC and voltage are chosen as the state variables:
	\begin{equation}
		s(t) = [V(t), \text{SoC}(t)]^{\top}.
	\end{equation}
	
	The action of the agent is the charge current value:
	\begin{equation}
		a(t) = I(t), \: { s.t. } \; \eqref{eq:I_constraint}.
	\end{equation}

	With the established voltage-to-SoH mapping and the TD3 algorithm described above, we define the reward function to address the optimization problem in \eqref{eq:problem_formulation}, achieving fast charging while minimizing battery degradation:
	\begin{equation}
		r(t) = r_{\text{SoC}}(t) + r_{\text{vol}}(t) + r_{\text{smooth}}(t). \label{eq:reward_all}
	\end{equation}
	
	To encourage fast charging, the SoC-related reward is defined as:
	\begin{equation}
		r_{\text{SoC}}(t) = \lambda_{\text{SoC}}\left| \text{SoC}^* - \text{SoC}(t) \right|,
	\end{equation}
	where $\lambda_{\text{SoC}}$ is a weighting factor that quantifies the importance of rapidly reaching the target SoC. To mitigate risks associated with lithium plating, the terminal voltage is constrained by the SoH-dependent upper bound, $V_{\text{max}}(\text{SoH})$:
	\begin{equation}
		r_{\text{vol}}(t) = 
		\begin{cases} 
			\lambda_{\text{vol}}\left| V(t) - V_{\text{max}}(\text{SoH}) \right|, & V(t) > V_{\text{max}}(\text{SoH}), \\
			0, & \text{otherwise},
		\end{cases}
	\end{equation}
	where $\lambda_{\text{vol}}$ is a weighting factor representing the importance of adhering to the voltage limit.
	
	Additionally, to encourage gradual variations in the charging current and prevent abrupt control actions that may stress the battery, we introduce a smoothness term:
	\begin{equation}
		r_{\text{smooth}}(t) = \lambda_{\text{smooth}}\left| a(t) - a(t-1) \right|,
	\end{equation}
	where $\lambda_{\text{smooth}}$ is a tuning parameter that penalizes rapid changes in the control input.

It is important to note that, despite the value of SoH being used to update the voltage constraints, it does not appear as a direct term in the reward function. Instead, the influence of SoH is embedded within the dynamic adjustment of $V_{\max}$ and the value of $V_{\max}$ ensuring that the reinforcement learning policy naturally adapts to the aging mechanism of battery without penalizing the agent explicitly for changes in SoH.

\section{Health-aware fast charging results}\label{sec:results}

In this section, we aim to evaluate the proposed RL-based charging strategy using high-fidelity simulations and compare it against widely adopted charging control strategies, such as CC-CV and CC-COP control. To evaluate the effectiveness of incorporating the $V_{\text{cutoff}}$–SoH mapping described in Section~\ref{sec:RL_method}, we enhanced the CC-CV method in which the voltage value during the constant-voltage stage is dynamically adjusted based on the SoH-dependent mapping. This modified approach is termed as the CC-CV with varying voltage (CC-CV-V) method, which aims to account for the SoH during battery charging. Additionally, we explore the constant current-constant overpotential method as a benchmark controller, setting the overpotential reference to two distinct values: $\eta_\text{side}^* = 0.01$ V for a conservative charging strategy, and $\eta_\text{side}^* = -0.05$ V for an aggressive approach. These variants are named CC-COP-slow and CC-COP-fast, respectively, and are used for comparative analysis.

The proposed approach does not require a physical model for implementation as long as the state and reward can be obtained. However, in the context of health-aware battery charging, conducting full-lifetime experiments to collect sufficient data is highly time-consuming. As the primary aim of this work is to introduce and validate an RL-based charging framework, we employ a high-fidelity single-particle model with electrolyte dynamics for training and performance demonstration. The details of the model are provided below.

\subsection{Battery environment for health-aware fast charging}\label{sec:model}
\subsubsection{Single particle model with electrolyte}
To provide a reliable and computationally efficient simulation environment for training and evaluating the proposed RL-based charging strategy, this work adopts the SPMe. While the DFN model, or called pseudo-two-dimensional (P2D) model, is a widely used and high-fidelity framework for simulating lithium-ion battery behavior \cite{xia2017computationally}, it involves a large number of parameters and intensive computational costs. Moreover, many of its parameters are difficult to measure and calibrate accurately, limiting its practicality in control-oriented applications \cite{khalik2021model}. In contrast, the SPMe offers a simplified yet sufficiently accurate representation by reducing both the parameter space and computational demand \cite{moura2016battery}. These features make the SPMe particularly suitable for simulating long-term battery performance and degradation, which are essential considerations in this study.

\begin{figure}
    \centering
    \includegraphics[width=0.65\linewidth]{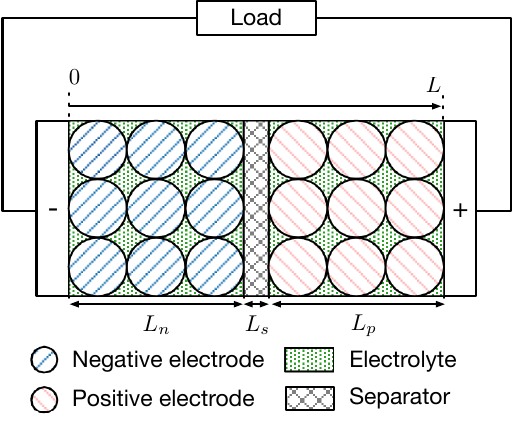}
    \caption{The internal structure of a lithium-ion battery cell (modified from \cite{marquis2019asymptotic}).}
    \label{fig:Li_battery}
\end{figure}
    
The structure of a lithium-ion battery is illustrated in Fig.~\ref{fig:Li_battery}, where the negative electrode, separator, and positive electrode have thicknesses \(L_n\), \(L_s\), and \(L_p\), respectively. The SPMe model approximates each electrode as a single spherical particle, assuming that the solid-phase concentration is uniformly distributed along the thickness direction, denoted by $x$. Under this assumption, the diffusion process inside the spherical particle is governed by
    \begin{equation}
		\frac{\partial c_{s,i}}{\partial t}(r,t) = \frac{D_{s,i}}{r^2}\frac{\partial}{\partial r}\left( r^2\frac{\partial c_{s,i}}{\partial r}(r,t) \right), \quad i \in \{n,p\}, \label{eq:cs}
    \end{equation}
where \( r \) represents the distance from the center of the spherical particle, $t$ is time, $c_{s,i}$ is the solid-phase lithium-ion concentration, and $D_{s,i}$ is its diffusivity. The electrolyte concentration in all domains are defined as $c_{e,n}$, $c_{e,\text{sep}}$, and $c_{e,p}$.

The electric potential in each electrode $i$ is given by 
    \begin{equation}
		\phi_{s,i} (x,t) = \eta_{i}(t) + \phi_{e,i}(x,t) + U_{i}\left( c_{ss,i}(t)\right) +F R_{f,i} j_{n,i} (t), %\quad i \in \{ n,p \},
    \end{equation}
where $c_{ss}$ is the solid-phase lithium-ion concentration on the particle surface, and $\eta_{i}(t)$ is the overpotential.

The nonlinear output of terminal voltage $V(t)$ is computed as: 
\begin{equation}
    V(t) = \phi_{s,p}(L,t) - \phi_{s,n}(0,t).
\end{equation}

% \textbf{Battery degradation mechanisms}
\subsubsection{Battery degradation mechanisms}
	
Battery performance and lifetime are limited by several degradation mechanisms \cite{wang2024onboard}. In this study, two dominant degradation processes, namely the SEI layer growth and lithium plating are considered. The idea is that suppressing them during fast charging may significantly extend battery lifetime.

The SEI layer forms on the anode particle surface and significantly affects the lifetime of lithium-based batteries. SEI growth is primarily a diffusion-limited process, in which solvent molecules react with lithiated graphite. To accurately capture this process, we utilize a two-layer diffusion-limited SEI growth model described in \cite{o2022lithium}. In this model, the inner and outer SEI layers are assumed to grow simultaneously at the same rate. 

Recent studies have indicated that degradation mechanisms, particularly lithium plating, are strongly influenced by the anode overpotential and can significantly limit battery performance and lifespan \cite{chaturvedi2010algorithms, tomaszewska2019lithium}. This anode overpotential is described as: 
	\begin{equation}
		\eta_{\text{side}}(x,t) = \phi_{s,n} - \phi_{e,n} - U_{\text{side}}, \label{eq:eta_sr}
	\end{equation}
where $U_{\text{side}}$ is the equilibrium potential of the side reaction, assumed to be zero for lithium plating \cite{arora1999mathematical}.

The initial partial differential-algebraic equation (PDAE) model that couples the SPMe and the aging model can be reformulated and reduced as a DAE model by suitable numerical methods, such as finite difference and spectral methods \cite{li2021model, huang2024minn}: 
	\begin{align}
		\dot{x} & = f(x,z,u), \\
		y & = h(x,z,u), \\
		0 & = g(x,z,u),
	\end{align}
where $x = [c_{s,n}, c_{s,p}, c_{e}]^{\top} \in \mathbb{R}^{n_{x}}$ is the state vector, $z = [\phi_{s,n}, \phi_{s,p}, \phi_{e}, i_{e,n}, i_{e,p} ]^{\top} \in \mathbb{R}^{n_z}$ is the algebraic variable vector, $y = [V, \text{SoC}, \text{SoH}]^{\top}$ is the output and $u \triangleq I$ is the applied input current. 

\subsection{Experiment and simulation configurations} % Simulation setup and training of the proposed method

The open-source PyBaMM software (version 24.5), is employed as the simulation platform \cite{sulzer2021python}, in which the SPMe-aging model from Section~\ref{sec:model} is implemented. As detailed in Section~\ref{sec:model}, the battery model features a single particle model with an electrolyte component, which facilitates efficient and comprehensive analysis of cycle aging. The investigated battery is an LG M50 cell, with electrochemical parameters obtained from \cite{o2022lithium}. The main degradation mechanisms are characterized by an SEI solvent diffusivity of $2.5 \times 10^{-22}$ m$^2$/s, a lithium plating kinetic rate constant of $1 \times 10^{-11}$ m/s, and an irreversible condition for lithium plating.

\begin{figure}[h!]
    \centering
    \begin{subfigure}[b]{0.8\linewidth}
	\includegraphics[width=\linewidth]{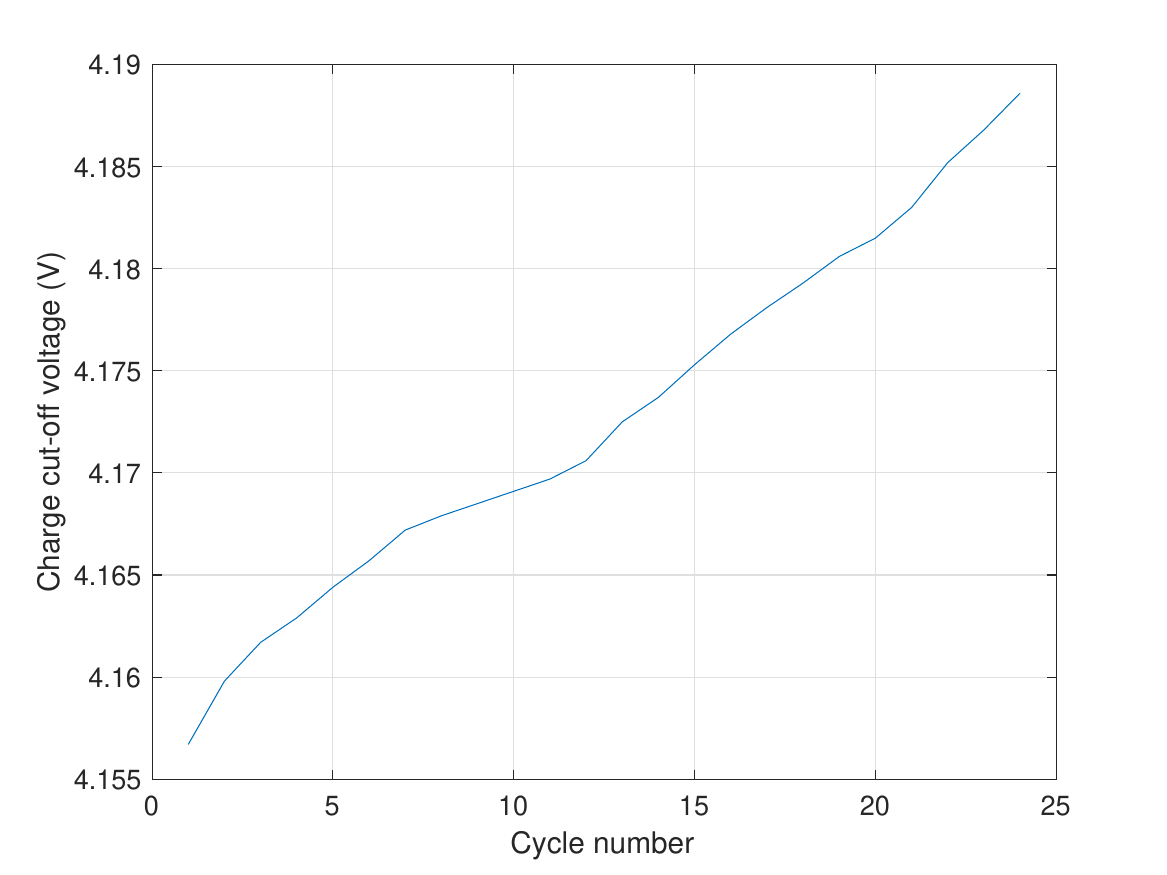}
	\caption{}\label{fig:vol_cutoff_exp}
    \end{subfigure}
    \hfill
    \begin{subfigure}[b]{0.8\linewidth}
        \includegraphics[width=\linewidth]{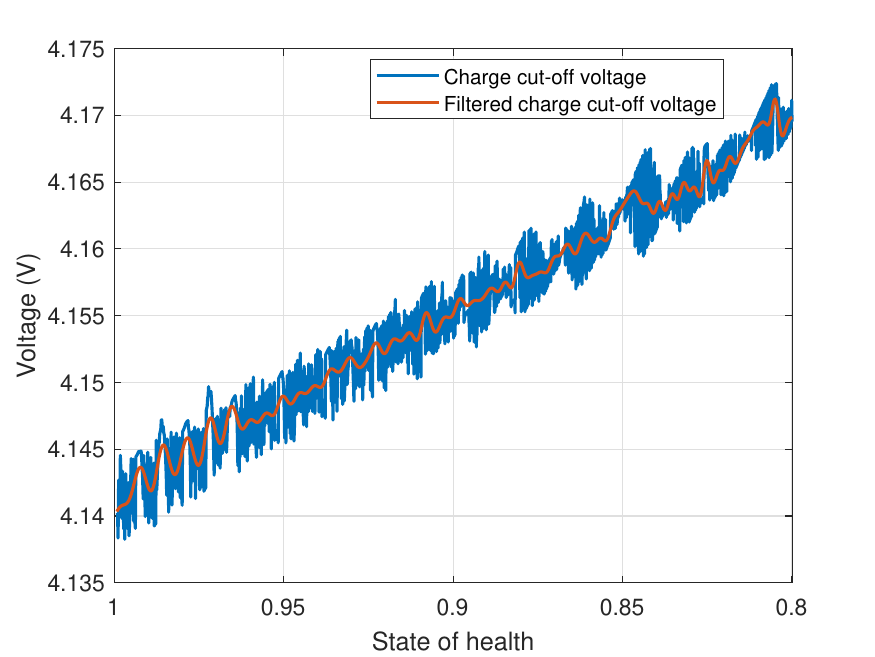}
	\caption{}\label{fig:voltage_end}
	\end{subfigure}
    \caption{Extraction and comparison of cut-off voltage–SoH mapping from (a) experimental data and (b) simulation.}\label{fig:cut-off-SOH}
\end{figure}

To obtain the mapping between cut-off voltage and SoH, we first configure the experimental setup as illustrated in Fig.\ref{fig:exp_setup}, and then compare the mappings derived from both experimental measurements and high-fidelity simulation, as shown in Fig.\ref{fig:cut-off-SOH}. Although different battery cells are used in the experiment and simulation, the trend between the charge cut-off voltage and SoH remains consistent across both domains.

For performance evaluation, we implement a unified framework to assess various charging strategies across the entire battery lifetime. The details of this process are shown as Protocol~\ref{alg:charging_step}. This protocol includes an initial stage to obtain the actual capacity of battery, followed by repeated testing cycles using different charging controls. The primary objective of this comprehensive battery life testing is to evaluate the effectiveness of these controls in optimizing the charging process and extending the operational lifespan of the battery.

\begin{algorithm}[h!]
    \caption{Battery cycling protocol for controller investigation }\label{alg:charging_step}
    \begin{algorithmic}[1]
		\State \textbf{Initialize:} Discharge the battery to 0\% SoC.
		\State \textbf{Charge} to 100\% SoC using CC-CV.
		\State \textbf{Discharge} the battery completely to determine actual capacity.
		\Repeat
		\State \textbf{Charge} the battery from 0\% SoC to 20\% SoC using low-current CC strategy.
		\State \textbf{Rest} for one hour.
		\State \textbf{Apply} the target charging strategy to charge the battery to 80\% SoC.
		\State \textbf{Discharge} the battery to 0\% SoC.
		\State \textbf{Rest} for one hour.
		\State \textbf{Check} SoH; if SoH falls to $\text{SoH}_{\text{end}}$ (e.g., 80\%), end test.
		\Until{battery SoH $\leq \text{SoH}_{\text{end}}$}
		\State \textbf{Evaluate} performance based on total cycles in terms of the averaged charging time and the equivalent full cycles.% (EFCs) or energy throughput..
    \end{algorithmic}
\end{algorithm}

\subsubsection{Controller configuration and training results}

The proposed controller employs an actor-critic neural network architecture. The actor and critic networks each contain two fully connected hidden layers with 200 units per layer. The hyperparameters of training are summarized in Table~\ref{tab:TD3_parameters}. The penalty coefficients in the proposed controller are chosen as $\lambda_{\text{SoC}} = -2$, $\lambda_{\text{vol}} = -10$, and $\lambda_{\text{smooth}} = -0.5$, respectively. The maximum control input is limited as $I_{\max} = 10$A.

	\begin{table}[h!]
		
		\caption{Hyperparameter of the proposed TD3 algorithm}\label{tab:TD3_parameters}
		
		\begin{centering}
			\begin{tabular}{cc}
				\hline \hline
				Hyperparameters & Values\tabularnewline
				\hline 
				Actor network learning rate & 0.0001\tabularnewline
				Critic network learning rate & 0.0001\tabularnewline
				Discounting factor & 0.99\tabularnewline
				Experience replay buffer size & $1.0 \times 10^{6}$ \tabularnewline
				Minibatch size & 256\tabularnewline
				Soft update factor & 0.005\tabularnewline
				Delay frequency & 1\tabularnewline
				\hline \hline
			\end{tabular}
			\par\end{centering}
	\end{table}
	
	\begin{figure}[h!]
		\centering
		\begin{subfigure}[b]{0.75\linewidth}
			\includegraphics[width=\linewidth]{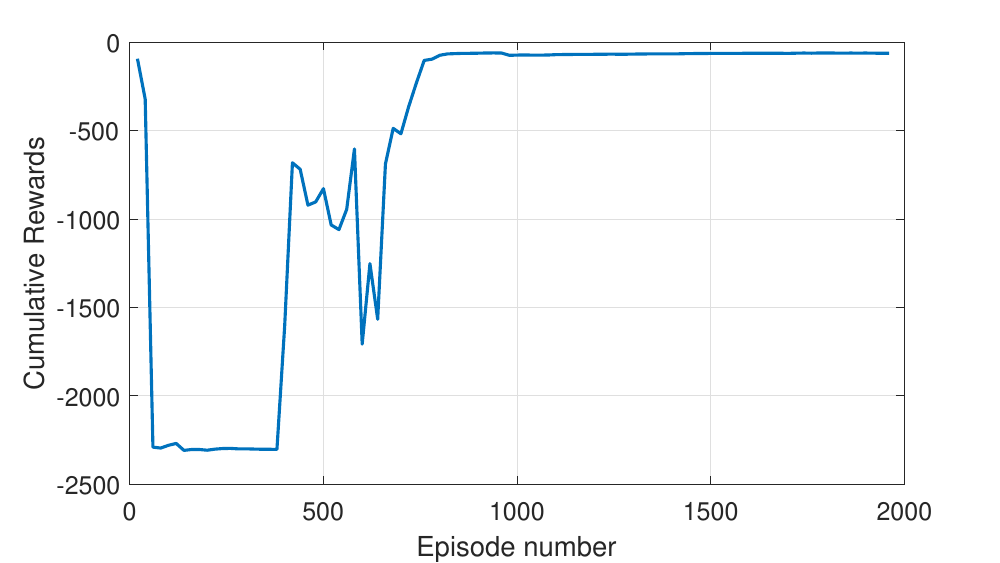}
			\caption{}\label{fig:RL_train_reward}
		\end{subfigure}
		\hfill
		\begin{subfigure}[b]{0.75\linewidth}
			\includegraphics[width=\linewidth]{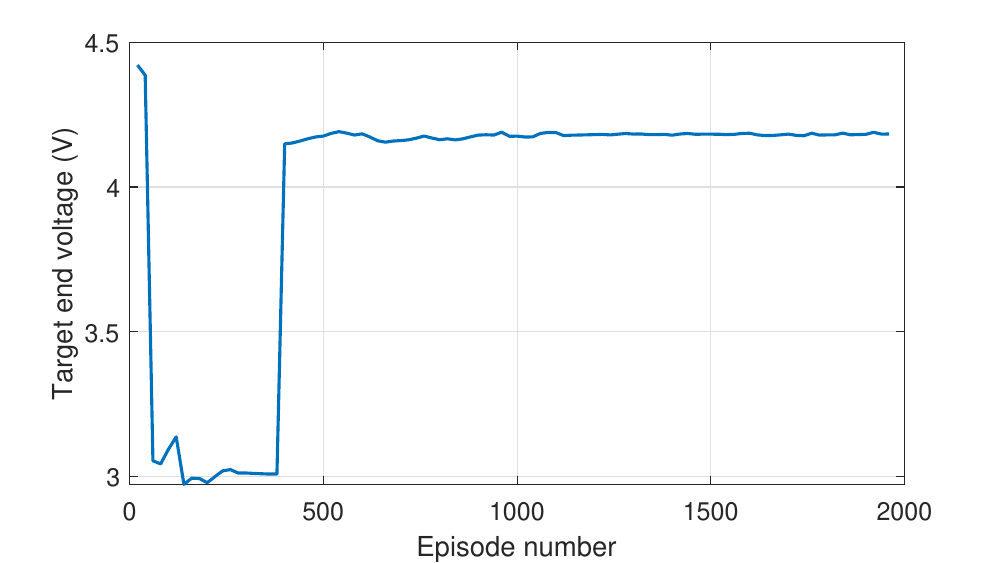}
			\caption{}\label{fig:RL_train_voltage}
		\end{subfigure}
		\begin{subfigure}[b]{0.75\linewidth}
			\includegraphics[width=\linewidth]{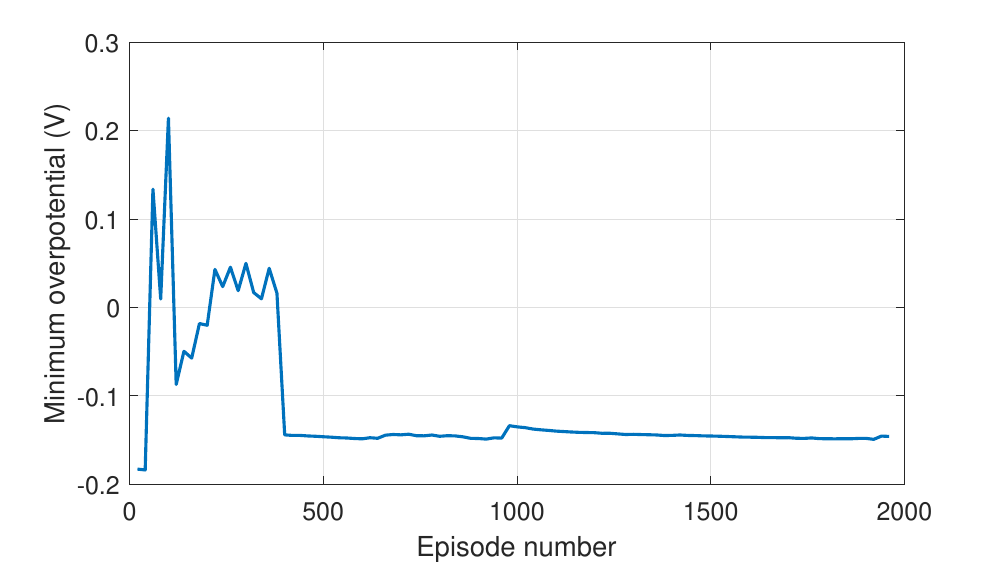}
			\caption{}\label{fig:RL_train_eta}
		\end{subfigure}
		\hfill
		\begin{subfigure}[b]{0.75\linewidth}
			\includegraphics[width=\linewidth]{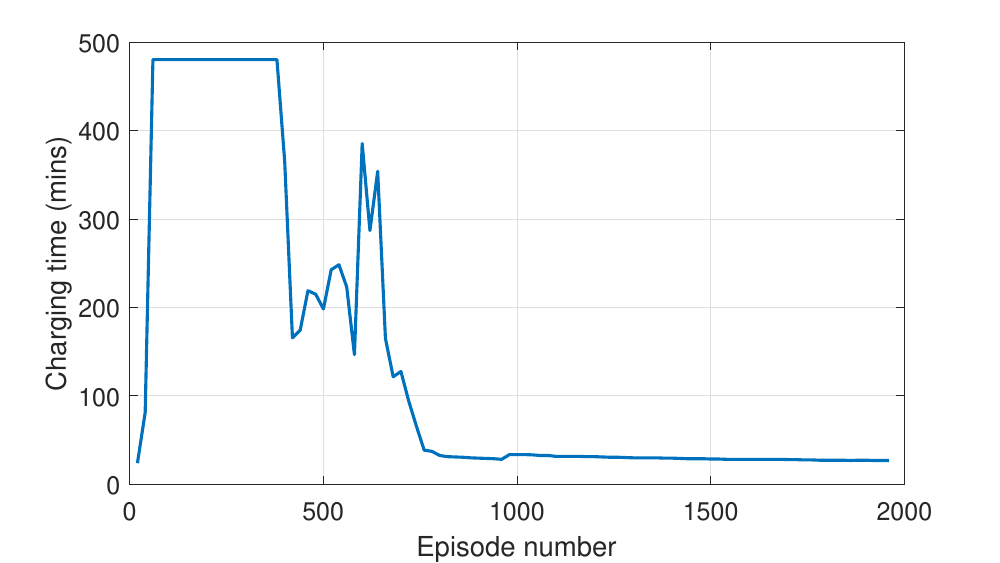}
			\caption{}\label{fig:RL_train_time}
		\end{subfigure}
		\caption{The training performance of the proposed method: (a) cumulative rewards; (b) maximum end voltage; (c) minimum overpotential $\eta_{\text{side}}$; (d) charging time.}\label{fig:RL_training}
	\end{figure}

The training performance of the proposed charging algorithm is presented in Fig.~\ref{fig:RL_training}, illustrating the evolution of four critical metrics across approximately 2000 episodes. The evaluations were conducted at 20 episode intervals. Fig.~\ref{fig:RL_train_reward} shows the cumulative reward, which initially fluctuates significantly around $-$2000 due to the exploration of highly conservative charging strategies but stabilizes near $-$60 after approximately 1000 episodes, indicating convergence to a policy effectively reducing charging time while extending battery life. The large initial negative values result from assigning a substantial negative reward ($-$1000) whenever the charging time exceeds 480 minutes. Fig.~\ref{fig:RL_train_voltage} depicts the charge cut-off voltage, which initially oscillates around 3.5 V but stabilizes around 4.2 V after 1000 episodes, demonstrating the ability of algorithm to maintain an optimal voltage threshold guided by the $V_{\text{cutoff}}$-SoH mapping. Additionally, Fig.~\ref{fig:RL_train_eta} illustrates the convergence behavior of the minimum overpotential, showing that its value stabilizes around $-$0.145 V. Fig.~\ref{fig:RL_train_time} demonstrates that, in general, higher cumulative rewards during training correlate with reduced charging times. The charging time ultimately stabilizes near 27 minutes in later episodes.

\subsection{Fast charging results}

\begin{table}[h!]
    \centering
    \caption{Battery charging performance based on different methods}\label{tab:EFCN_time}
    \begin{threeparttable}
        \begin{tabular}{ccc}
            \hline \hline
            & Maximum EFCs & Average charging time (mins) \tabularnewline
            \hline 
            CC-CV & 572 & 24.15\tabularnewline
            CC-CV-V & 611 & 24.14\tabularnewline
            CC-COP-slow\tnote{*} & 1316 & 36.02\tabularnewline
            CC-COP-fast\tnote{*} & 1009 & 22.40\tabularnewline
            Proposed method & 703 & 24.12\tabularnewline
            \hline \hline
        \end{tabular}
        \begin{tablenotes}
            \small 
            \item[*] Requires side-reaction overpotential measurements.
        \end{tablenotes}
    \end{threeparttable}
\end{table}
	
Following the battery cycling protocol as described in Protocol~\ref{alg:charging_step}, we conducted the battery charging tests using five different strategies: the CC-CV strategy, the CC-CV-V strategy, two CC-COP strategies, and the proposed approach. A finer sampling resolution can better capture the transient dynamics of batteries, but at the expense of computational efficiency. In this study, the primary metrics of interest are battery lifetime and charging time. Considering these factors, the sampling time for the lifelong implementation of all these strategies was set as $20$ s. The EFCs and the average charging time for these approaches are summarized in Table~\ref{tab:EFCN_time}. %

\subsubsection{Overall analysis on charging time and lifespan}
Among the methods, the CC-COP controllers rely fully on the measurement of side-reaction overpotential, and demonstrate performance variation depending on the tuning of the desired overpotential reference. The CC-COP-fast strategy achieves the shortest average charging time of 22.40 minutes and significantly extends the battery lifetime compared to CC-CV methods, with 1009 EFCs. In comparison, the CC-COP-slow strategy attains the longest battery lifespan, which is 1316 FECs, yet requires the longest charging time of 36.02 minutes, reflecting a very conservative strategy. 

\begin{figure}[h!]
    \centering
    \includegraphics[width=0.8\linewidth]{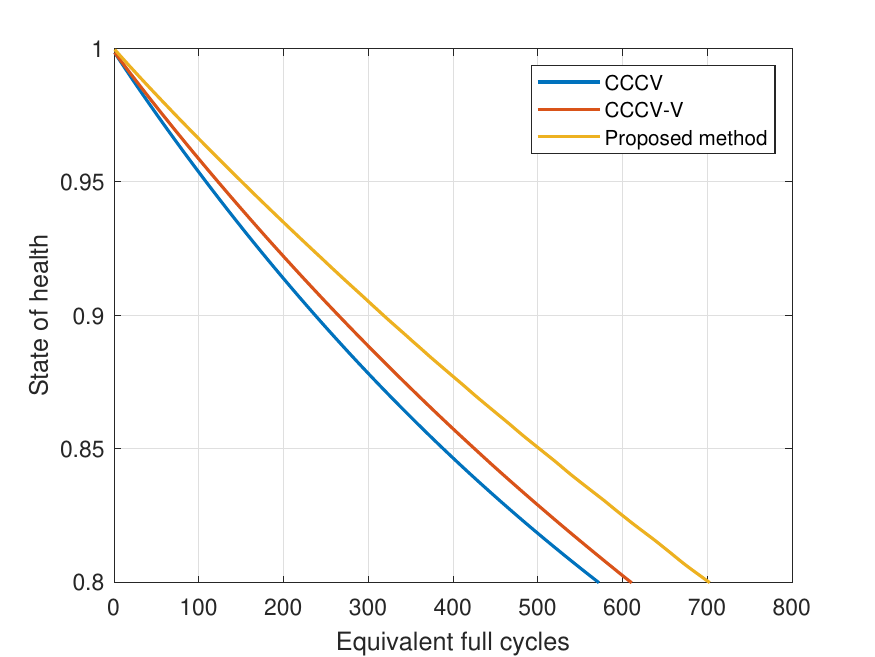}
    \caption{Comparison of SoH degradation among different charging methods.}
    \label{fig:SoH_vs_EFCs}
\end{figure}

Different from CC-COP strategies, the proposed approach along with CC-CV and CC-CV-V exhibit more practical applicability and deployability. The variation in battery SOH with respect to EFCs for these three methods is presented in Fig.~\ref{fig:SoH_vs_EFCs}. The baseline CC-CV method, despite achieving a competitive charging time of 24.15 minutes, suffers from the shortest battery lifespan with 572 EFCs. This limitation stems from its rigid voltage and current thresholds during the charging process, which fail to adapt to battery degradation dynamics. The CC-CV-V strategy addresses this weakness by incorporating $V_{\text{cutoff}}$-SOH mapping to dynamically adjust the charge cut-off voltage. This modification yields a modest lifespan improvement with 611 EFCs while maintaining near-identical charging speed with 24.14 minutes. However, the 6.8\% improvement of EFCs over CC-CV suggests that static parameter mapping alone cannot fully capture complex aging mechanisms. 

In contrast, the proposed method achieves a significant improvement in performance, where battery lifespan is extended to 703 EFCs (23\% and 15\% higher than CC-CV and CC-CV-V, respectively) while marginally reducing charging time to 24.12 minutes. This breakthrough results from its health-aware control architecture, which integrates indirect degradation indicators to dynamically optimize charging protocols. %

In summary, compared to CC-COP strategies, the proposed method achieves enhanced practicality for real-world deployment by eliminating the need for specialized overpotential sensors, despite its slightly longer charging time and fewer equivalent full cycles than CC-COP-fast and CC-COP-slow approaches. In contrast to CC-CV and its variants, the proposed approach maintains comparable charging efficiency while largely extending battery lifespan, % 
demonstrating that lifespan enhancement can be achieved without compromising charging speed.

\begin{figure}[h!]
    \centering
    \includegraphics[width=0.75\linewidth]{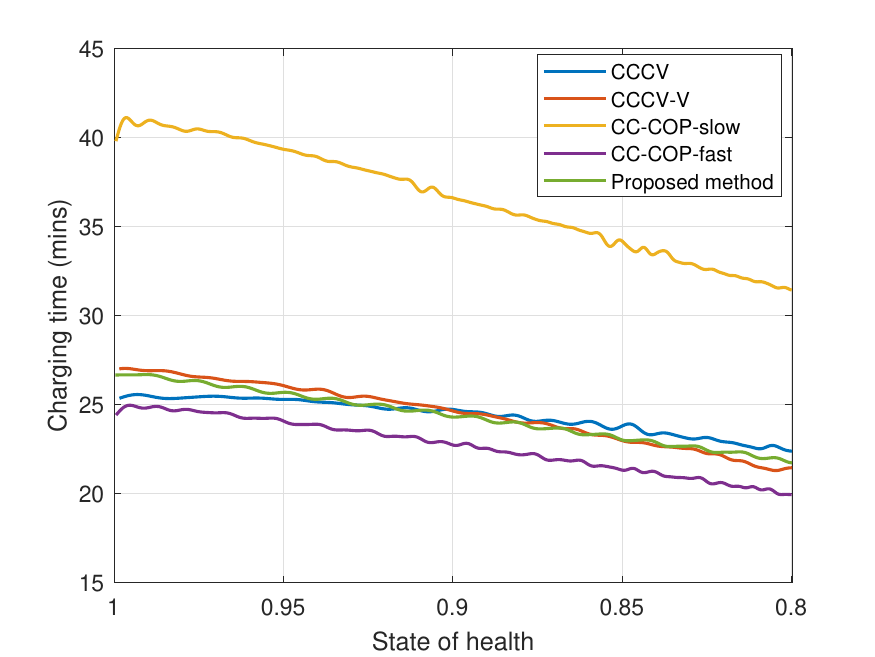}
    \caption{Charging time based on different charging approaches.}
    \label{fig:time_vs_SoH}
\end{figure}

\subsubsection{Charging time dynamics over battery aging}
Fig.~\ref{fig:time_vs_SoH} compares five charging strategies in terms of charging time versus SoH. Overall, the CC-COP-slow method exhibits the longest charging durations across the SoH range, starting at approximately 40 minutes when the battery is fully healthy and dropping to around 30 minutes when SoH reaches 80\%. The charging times of all methods decrease as SoH declines, primarily because less energy or charge capacity is absorbed by the battery for the same SoC change. The CC-CV, CC-CV-V, and proposed methods exhibit generally similar trajectories, though CC-CV-V notably shows a substantial peak at the beginning and subsequently decreases more rapidly than the CC-CV and the proposed method. By contrast, the proposed method maintains a relatively stable decrease in charging time compared to the other methods.

\subsubsection{Evolution of current and overpotential profiles}
\begin{figure}[h!]
		\centering
		\begin{subfigure}[b]{0.75\linewidth}
			\includegraphics[width=\linewidth]{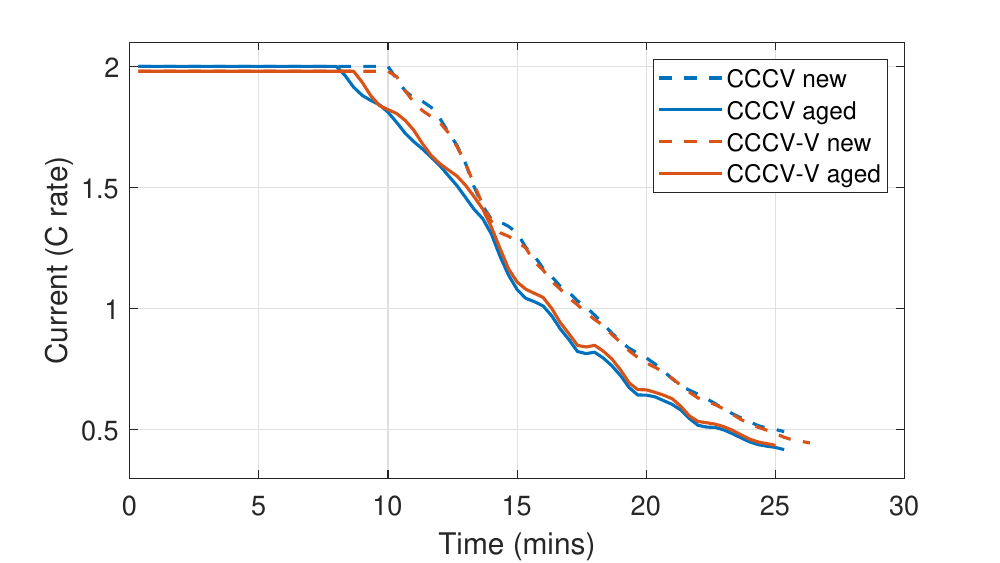}
			\caption{}\label{fig:CC-CVs_current_cycle}
		\end{subfigure}
		\hfill
		\begin{subfigure}[b]{0.75\linewidth}
			\includegraphics[width=\linewidth]{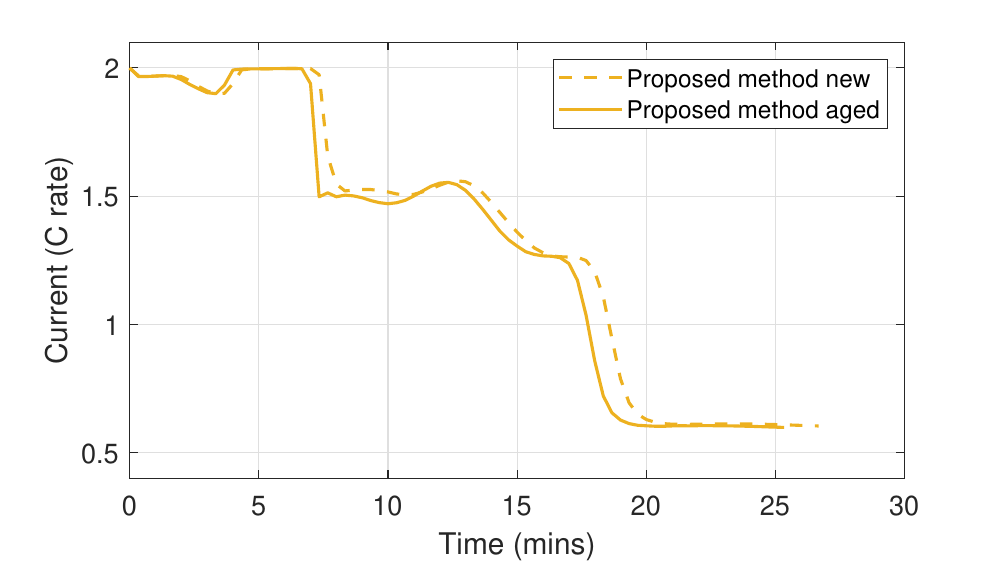}
			\caption{}\label{fig:RL_voltage}
		\end{subfigure}
		\begin{subfigure}[b]{0.75\linewidth}
			\includegraphics[width=\linewidth]{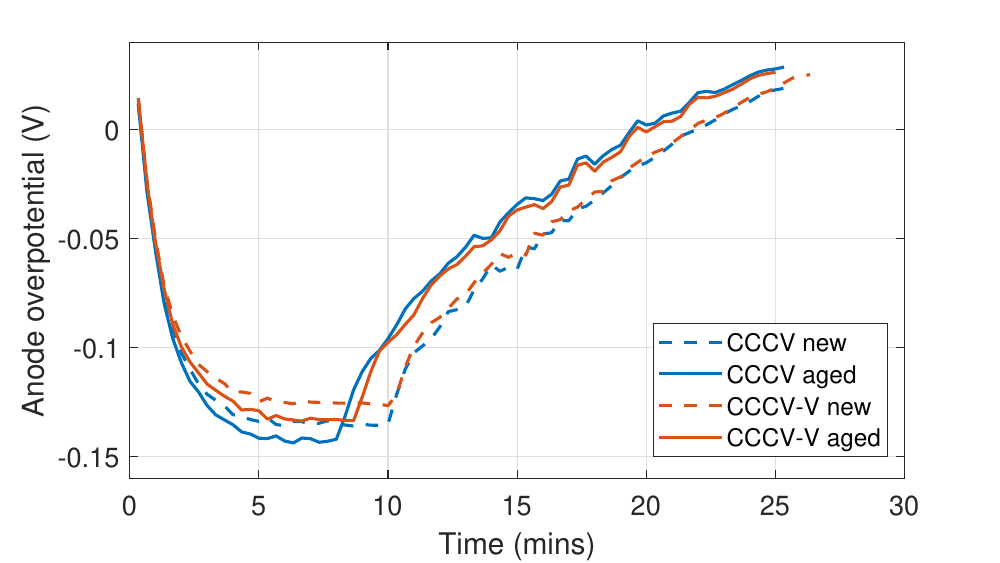}
			\caption{}\label{fig:CC-CVs_eta_cycle}
		\end{subfigure}
		\hfill
		\begin{subfigure}[b]{0.75\linewidth}
			\includegraphics[width=\linewidth]{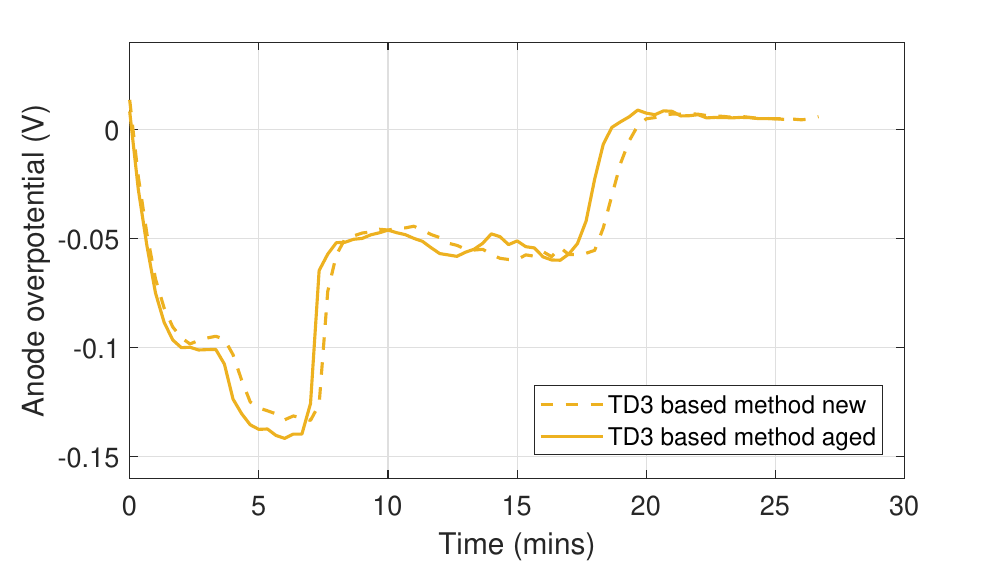}
			\caption{}
		\end{subfigure}
		\caption{Current and anode potential profiles of the studied cell at different aging levels under different control methods.}\label{fig:result_cycles}
	\end{figure}

To further investigate why the proposed learning-based method effectively extends the battery lifespan compared to CC-CV and its variants, the battery current and anode overpotential profiles during a single charging process are plotted for two distinct aging stages. Specifically, results from the 1st cycle (new battery) and the 200th cycle (aged battery) are presented in Fig.~\ref{fig:result_cycles}.

Fig.~\ref{fig:result_cycles} illustrates distinct shifts in anode overpotential profiles across different charging methods when comparing new and aged cells. For aged cells, a clear downward and leftward shift in the overpotential curves is observed, indicating that the same charging current results in lower overpotential as the cell ages. This phenomenon is particularly pronounced in the CC-CV and CC-CV-V charging methods. In contrast, the current profiles of the proposed method behave like a multi-stage constant-current scheme that results in smoother current and anode potential transitions in both new and aged cells. The TD3-based method significantly reduces the difference in anode potentials between new and aged cells, indicating its ability to attenuate the effects of battery aging. %

Regarding overpotential levels, the proposed approach only remains at a relatively low negative value for a short period before moving the anode potential closer to zero. By contrast, the overpotential under CCCV and its variants methods remains below $-$0.05 V for most of the charge and lasts for nearly 15 minutes. This prolonged negative overpotential may thermodynamically favor lithium plating, thereby increasing the risk of degradation and reducing long-term battery health compared to the proposed method. %
	
These results not only facilitate a comparison between the different controllers but also highlight the discrepancies in performance of the same controller at different stages of battery aging. Such insights are especially meaningful for researchers and engineers seeking to optimize fast charging strategies. %

\subsubsection{Capacity loss due to lithium plating}
\begin{figure}[h!]
		\centering
		\includegraphics[width=0.7\linewidth]{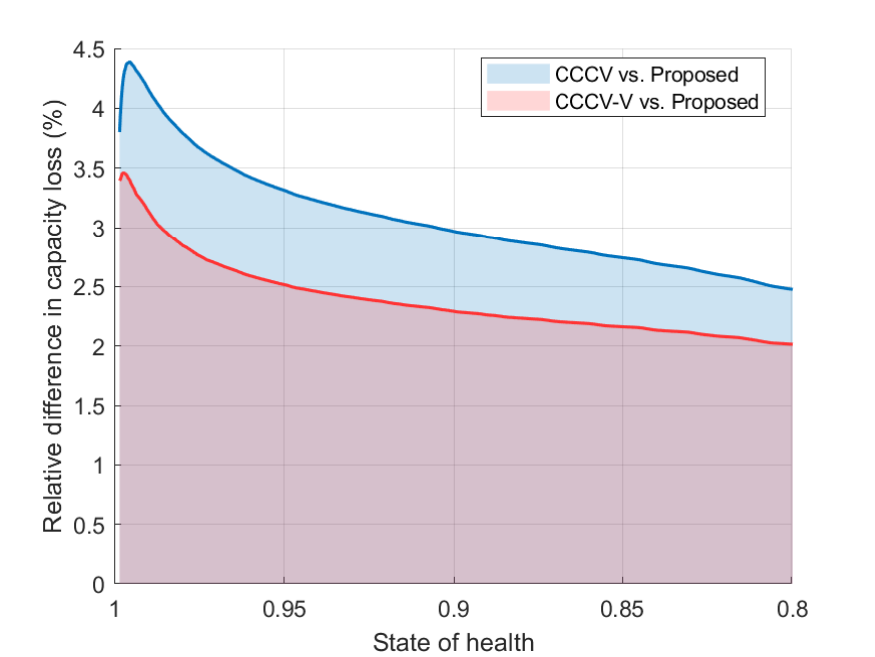}
		\caption{Relative difference in capacity loss due to lithium plating by comparing CC-CV and CC-CV‐V to the proposed method. The positive value in $y$-axis means the least plated lithium triggered by the proposed method.}
		\label{fig:cap_loss_Li}
	\end{figure}

Finally, we analyze the relative difference in capacity loss for CC-CV and CC-CV-V, compared to the proposed method, as a function of SoH, as shown in Fig.~\ref{fig:cap_loss_Li}. This difference is expressed as a percentage of the baseline capacity loss for each strategy. Positive values indicate that the proposed method yields a lower capacity loss than conventional strategies, highlighting its effectiveness in extending battery life.

\section{Conclusion}\label{sec:conclusion}
	
In this paper, we present a novel health-aware fast charging strategy for lithium-ion batteries, based on a deep learning approach. Unlike conventional charging methods that primarily focus on minimizing charging time, our proposed approach explicitly considers the trade-off between fast charging and extending battery life. By leveraging a mapping between the charge end voltage and the SoH of the battery from a constant current and constant overpotential control, the proposed method incorporates SoH-dependent voltage into the optimal control decision-making process, effectively mitigating adverse degradation phenomena. To demonstrate the effectiveness of the proposed method, we utilized a high-fidelity single particle model with electrolyte implemented in PyBaMM. This model served as the test environment, capturing realistic degradation behaviors and allowing for robust evaluation across a full life-cycle simulation. Comparisons with benchmark controllers demonstrate that the TD3-based policy successfully reduces overall degradation without significantly compromising charge speed, in a practical and implementable way.

\bibliographystyle{IEEEtran}

\bibliography{references.bib}

\vfill

\end{document}